\documentclass{amsart}

\usepackage{latexsym}
\usepackage{amssymb}
\usepackage{amsmath}

\theoremstyle{plain}
\newtheorem{theorem}{Theorem}[section]
\newtheorem{lemma}{Lemma}[section]
\newtheorem{prop}{Proposition}[section]

\newtheorem*{rmk}{Remark}

\def\R{\mathbb R}
\def\Om{\Omega}
\def\Sf{\Sigma}

\def\Uo{U^{2\epsilon}_{+}}
\def\Ui{U^{2\epsilon}_{-}}

\def\Sfsig{\Sigma_\sigma}
\def\Sfsigz{\Sigma_{\sigma_0}}
\def\tM{\tilde{M}}
\def\tMinf{\tilde{M}_\infty}
\def\tMsinf{\tilde{M}^s_\infty}
\def\tMkinf{\tilde{M}^{s_k}_\infty}
\def\Minf{M_\infty}

\def\PM{\mathcal{PM}}
\def\lPM{\overline{\PM}}
\def\rto{\rightarrow}
\def\tH{\tilde{H}}

\def\go{g_{+}}
\def\gos{{g_s}_{+}}
\def\got{{g_t}_{+}}

\def\gs{g_s}

\def\gi{g_{-}}

\def\gsc{g^s_c}
\def\gdel{g_{\delta}}
\def\gsdel{g^s_\delta}

\def\Rsdel{R^s_\delta}

\def\tR{\tilde{R}}

\def\tRsdel{\tilde{R}^s_\delta}
\def\tRkdel{\tilde{R}^{s_k}_{\delta_k}}

\def\usdel{u^s_\delta}

\def\vsdel{v^s_\delta}
\def\vkdel{v^{s_k}_{\delta_k}}

\def\tg{\tilde{g}}
\def\hg{\hat{g}}

\def\tgsdel{\tilde{g}^s_\delta}
\def\tgkdel{\tilde{g}^{s_k}_{\delta_k}}

\def\hgsdeltz{{\hg}^s_{\delta t_0}}
\def\hgsdel{{\hg}^s_\delta}

\def\Mb{\mathcal{M}_\infty}

\def\a{\alpha}

\def\ep{\epsilon}
\def\del{\delta}
\def\sig{\sigma}

\def\pdt{\frac{\partial}{\partial t}}

\def\vn{\vec{n}}
\def\pvn{\partial \vn}

\def\vn{\vec{n}}

\def\norm{||}
\def\lbar{\overline}
\def\stop{\hfill$\Box$}
\def\pf{\noindent \emph{Proof: }}

\def\deltoz{\lim_{\del \rightarrow 0}}

\def\g0{\go(0)}

\def\12{\frac{1}{2}}

\def\ep{\epsilon}

\def\del{\delta}

\def\sig{\sigma}

\begin{document}

\title[Variational Effect of Boundary Geometry on ADM Mass]
{Variational Effect of Boundary Mean Curvature on ADM Mass
in General Relativity}

\author{Pengzi Miao}

\address{Mathematical Sciences Research Institute, Berkeley, CA 94720, USA}

\email{pengzim@msri.org}

\date{August, 2003}

\begin{abstract}
{We extend the idea and techniques in \cite{Miao} to study variational 
effect of the boundary geometry on the ADM mass of an asymptotically flat 
manifold. We show that, for a Lipschitz asymptotically flat metric 
extension of a bounded Riemannian domain with quasi-convex boundary,
if the boundary mean curvature of the extension is dominated by but not identically equal 
to the one determined by the given domain, we can decrease its ADM mass while 
raising its boundary mean curvature. Thus our analysis implies that, for 
a domain with quasi-convex boundary, the geometric boundary condition holds in 
Bartnik's minimal mass extension conjecture \cite{Bartnik_energy}.}
\end{abstract}

\maketitle

\section{Introduction} \label{Intro}
Asymptotically flat manifolds are often used to model isolated systems
in general relativity.
A complete Riemannian manifold $(M^n, g)$ with dimension $n \geq 3$ is called 
{\bf asymptotically flat} if there is a compact set $K \subset M$ and a  
diffeomorphism 
$$\Phi: M \setminus K \rightarrow \R^n \setminus \{ |x| < 1 \} $$
 such that, in the coordinate chart defined by
$\Phi$, 
$$ | g_{ij}(x) - \del_{ij} | +  |x| | g_{ij,k}(x) | + 
|x|^2 | g_{ij,kl}(x)| = O(|x|^{-p}) $$
and $$ |R(g)(x)| = O(|x|^{-q}) $$
for some $p > \frac{n-2}{2}$ and some $q > n$, where ``;'' 
denotes partial derivative in the coordinate chart and $R(g)$ denotes 
the scalar curvature of $(M^n, g)$. 
The metric decay assumptions imply the existence of the limit
$$ m(g) = \frac{1}{ 4 \omega_{n-1} } \lim_{r \rightarrow \infty} 
{\oint}_{ |x| = r} \sum_{i,j}(g_{ij,i} - g_{ii,j} ) \nu^j d \mu , $$ 
where $\omega_{n-1}$ is the volume of the standard unit sphere
$\mathbb{S}^{n-1}$, $d \mu$ is the Euclidean surface measure and 
$\nu^j$ denotes the Euclidean unit normal. The quantity $m(g)$ 
is called the {\bf {\em total mass}} or {\bf{\em ADM mass}} 
of $(M^n, g)$ \cite{ADM}. It is a 
simple computation to show that if the metric $g$ is conformally flat 
and scalar flat, then the total mass appears in
the expansion of the conformal factor at infinity 
$$ u = 1 + \frac{A}{|x|^{n-2}} + O(|x|^{1-n}) $$
as $m(g) = (n-1)A$.

A fundamental result relating the total mass of an asymptotically flat 
manifold and its local energy density(scalar curvature) is the  
Positive Mass Theorem (PMT), first proved by R. Schoen and S.T. Yau 
\cite{Sch-Yau} using minimal surface techniques and later by E. Witten 
\cite{Witten} using spinors. 

\vspace{.3cm}

\noindent {\bf Positive Mass Theorem} \\
{\em Let $(M^n, g)$ be asymptotically flat with $R(g) \geq 0$. If $n \leq 7$ 
or $M$ is spin, then the total mass of $(M^n, g)$ is 
non-negative, and is zero if and only if $(M^n, g)$ is isometric to the 
Euclidean space $(\R^n , g_0)$.}

\vspace{.3cm}

Many other significant works have been made in the last two decades to 
understand the interplay between the total mass of $(M^n, g)$ and 
its geometry. Among them, one remarkable result is the following Riemannian 
Penrose Inequality proved by  H. Bray \cite{Bray-pen} and 
G. Huisken and T. Ilmanen 
\cite{oneoverH}.

\vspace{.3cm}

\noindent {\bf Riemannian Penrose Inequality} \\
{\em Let $(M^3, g)$ be asymptotically flat with $R(g) \geq 0$. 
Let $A$ be the area of the outermost minimal surface $\Sf$ in $(M^3, g)$.
Then 
$$ m(g) \geq \sqrt{\frac{A}{16\pi}} ,$$
and the equality holds if and only if the part of $(M^3, g)$ outside $\Sf$
is isometric to the Schwarzschild manifold 
$(\R^3 \setminus B_\frac{m}{2}(0), (1+ \frac{m}{2|x|})^4 g_0)$ with 
$m = m(g)$. }

\vspace{.3cm}

One natural question coming from the Penrose Inequality is, given an 
asymptotically flat $(M^n ,g)$, what the least contribution of a finite 
region $\Om \subset M$ to the total mass $m(g)$ is?  
Another way of asking the question is, between the notion of local
energy density and the notion of the total mass, if there is a meaningful 
concept of the mass of a bounded region?
There have been many attempts to define such a quasi-local mass 
function(\cite{Bartnik_local}, 
\cite{York-Brown}, \cite{Yau-Liu} etc.), and one believes there should be an 
analog in Einstein's gravity theory of the usual Newtonian measure of the 
mass of an extended body. 
In \cite{Bartnik_local}, R. Bartnik gave his quasi-local mass definition
$m_B({\Om})$ from a variational point of view,
$$ m_B(\Om) = \inf \{ m(\tg) \ | \  (\tM, \tg) \in \PM \}, $$
where 
\begin{eqnarray*}
\PM & = & \{ (\tM, \tg) \  | \ (\tM, \tg) \ is\ asymptotically\ flat\ with\ 
R(\tg) \geq 0, \\
& & \hspace{1.5cm}  (\tM, \tg)\ contains\ (\Om, g)\ isometrically, \\
& & \hspace{1.5cm} and\ no\ horizon\ lies\ outside\ (\Om, g). \}. 
\end{eqnarray*}
It has been shown in \cite{oneoverH} that $m_B(\Om) = 0$ if and only 
if $(\Om, g)$ is locally Euclidean and 
$\lim_{i \rightarrow \infty} m_{B}(\Om_i) = m(g)$
if $\{ \Om_i \}_{i=1}^\infty $ forms an exhaustion sequence of $(M, g)$.

There is a natural analogue between $m_B(\Om)$ and the usual 
definition of the electrostatic capacity of a conducting body,
$$ c(\Om) = \inf \left\{ \int |\nabla u|^2 dx \ |  \ u \in C^\infty_c(\R^3), \ 
u \equiv 1 \ on\ \Om \right\} , $$
where $c(\Om)$ is achieved by a harmonic function $u$ on $\R^3 \setminus 
\lbar{\Om}$ that equals $1$ on $\partial \Om$ and decays to $0$ at infinity.
It is interesting to know if similar things hold for $m_B(\Om)$, i.e. if 
there exists a metric $\tg$ on $M \setminus \Om$ such that 
$m(\tg) = m_B(\Om)$, and if it 
exists, what kind of interior equation and boundary condition it satisfy?

Both of the research in \cite{Miao} and in this paper are inspired by 
the above variational approach to the quasi-local mass problem. 
Motivated by the expectation that a metric achieving $m_B(\Om)$ 
might only be $Lipschitz$ across $\partial \Om$,
we established the positivity of the total mass of 
a class of piecewise smooth asymptotically flat manifolds containing
$(\Om, g)$ in \cite{Miao}. 
In this paper, we focus on the variational
effect of the boundary mean curvature on the total mass and relate it
to the geometric boundary condition in Bartnik's minimal mass 
extension conjecture \cite{Bartnik_energy}. 

I want to thank professor Richard Schoen for many helpful discussions.

\section{The Mass of Piecewise Smooth Manifolds} \label{Mass}
We first recall notations and results in \cite{Miao}.
Let $M^n$ be a differentiable manifold which has 
the property that there exists a compact domain $\Om$ with smooth boundary 
such that $M \setminus \Omega$ is diffeomorphic to $\R^n$ minus a ball.
Let $n \geq 3$ be a dimension for which the classical 
PMT \cite{Sch-Yau} holds.

\begin{theorem} \cite{Miao} \label{PMT}
Let $\gi$ and $\go$ be smooth metrics defined on $\lbar{\Om}$ and 
$M \setminus \Om$ so that $\gi |_{\partial \Om} = \go |_{\partial \Om}$ 
and $\go$ is asymptotically flat.
Suppose that both $\gi$ and $\go$ have non-negative scalar curvature and
$$ H(\partial \Om, \gi) \geq H(\partial \Om, \go) , $$
where $H(\partial \Om, \gi)$ and $H(\partial \Om, \go)$ represent the mean 
curvature of $\partial \Om$ in $(\lbar{\Omega}, \gi)$ and $(M \setminus 
\Omega, \go)$ with respect to unit normal vectors pointing to the 
unbounded region.

Then the mass of $\go$ is non-negative. If 
$H(\partial \Om, \gi) > H(\partial \Om, \go)$ 
at some point on $\partial \Om$, then $\go$ has a strict positive mass. 
If $n=3$ and the mass of $\go$ is zero,  
then $(\lbar{\Om}, \gi)$ can be isometrically embedded in $\R^3$ and 
$(M \setminus \Om, \gi)$ is isometric to its complement. 
\end{theorem}

\begin{rmk}
Our sign convention for the mean curvature is that $H(S^{n-1}, g_0)=n-1$,
where $S^{n-1}$ is the unit sphere in the Euclidean space $(\R^n, g_0)$.
\end{rmk}

The proof of this theorem in \cite{Miao} was based on Schoen-Yau's original 
proof of the classical PMT and a metric mollification proposition which 
interpretates the difference of the mean curvature as scalar curvature 
concentration along
the boundary. To state that proposition precisely, we 
let $U_{-}^{2\ep}$ and $\Uo$ be $2\ep$-tubular 
neighborhoods of $\Sf$ in $(\lbar{\Om}, \gi)$ and $(M \setminus \Om, \go)$ 
such that $\Ui$ and $\Uo$ are diffeomorphic to $\Sf \times (-2\ep, 0]$ and 
$\Sf \times [0, 2\ep)$, and $\gi |_{\Ui}$ and $\go |_{\Uo}$ have the form
\begin{eqnarray} 
\gi = {g_{-}}_{ij}(x,t)dx^idx^j + dt^2 \\
\go = {g_{+}}_{ij}(x,t)dx^idx^j + dt^2
\end{eqnarray}
where $t$ is the standard coordinate for $(-2\ep, 0]$ and $[0, 2\ep)$, and 
$(x^1, \ldots, x^{n-1})$ are local coordinates for $\Sf$. 
Identifying $U = \Ui \cup \Uo$ with $\Sf \times (-2\ep, 2\ep)$, we define 
$\tM$ to be a possibly new differentiable manifold with the background 
topological space $M$ and the differential structure determined by the open 
covering $\{\Omega, M \setminus \lbar{\Omega}, U \}$, where $U$ carries the 
differential structure induced from $\Sf \times (-2\ep, 2\ep)$. It follows 
from the fact 
$g_{-}|_\Sf = g_{+}|_\Sf$ that $\gi$ and $\go$ determines a continuous metric 
$g$ on $\tM$ such that $g |_{U}$ has the form
\begin{equation} 
 g = g_{ij}(x,t)dx^idx^j + dt^2, 
\end{equation}
where $ g_{ij}(x,t) = {g_{-}}_{ij}(x,t) $ when $ t\leq 0 $ and 
$ g_{ij}(x,t) = {g_{+}}_{ij}(x,t) $ when $ t\geq 0 $. For such a metric $g$, 
we have the following proposition
 
\begin{prop} \cite{Miao} \label{smthcornerapp}
There exists a family of $C^2$ metrics 
$\{ \gdel \}_{0 < \del \leq \ep}$ on $\tM$ such that $\gdel$ agrees 
with $g$ outside $\Sf \times (-\frac{\del}{2}, \frac{\del}{2})$, $g_\del$ 
is uniformly close to $g$ in $C^0$ topology and 
the scalar curvature of $\gdel$ satisfies 
\begin{eqnarray}
  R_\del(x,t) &  = & O(1), \ \ for\  (x,t) \in \Sf \times \{\frac{\del^2}{100}
  < |t| \leq \frac{\del}{2}\} \\
  R_\del(x,t) &  = & O(1) + \{H(\Sf, \gi)(x) - H(\Sf, \go)(x) \} \left\{ 
  \frac{100}{\del^2}\phi(\frac{100t}{\del^2}) \right\}, \nonumber \\
  & & for \  (x,t) \in \Sf \times [-\frac{\del^2}{100},\frac{\del^2}{100}] ,
\end{eqnarray}
where $O(1)$ represents bounded quantity depending only 
on $g$, but not on $\del$.
\end{prop}

\section{Variational Effect of Boundary Geometry} \label{bdryrv}
The main goal of this paper is to investigate 
the boundary mean curvature equality $H(\Sf, \gi) \equiv H(\Sf, \go)$ 
from a variational point of view.
We will briefly discuss its implication to the quasi-local mass question in 
the end.

Let $(M^n, g)$ be a smooth asymptotically flat manifold with 
$R(g) \geq 0$. Let $\Om \subset M$ be a compact domain with smooth 
boundary $\Sf$. We define
\begin{eqnarray} 
 \mathcal{M}_\infty & = & \{ (\Minf, \go) \ | \ 
(\Minf, \go) \ is\ a\ smooth\ asymptotically\ flat\ manifold \nonumber \\
 & & \hspace{2cm} with\ boundary\ \Sf \ such\ that\ R(\go) \geq 0, 
\nonumber \\
& & \hspace{2cm} g|_{\Sf} = \go |_{\Sf} \ and\
 \ H(\Sf, g) \geq H(\Sf, \go). \} \nonumber
\end{eqnarray}
Our next theorem states that, if $\partial \Om$ is quasi-convex in the sense
of \cite{Weinstein} and 
if $(\Minf, \go) \in \Mb$ and $H(\Sf, \go)$ does not agree with 
$H(\Sf, g)$ identically, we can decease $m(\go)$ while raising 
$H(\Sf, \go)$ to be almost $H(\Sf, g)$. 

\begin{theorem} \label{thdm}
Assume that $\Sf$ has positive scalar curvature with respect to the induced
metric and $H(\Sf, g) > 0$. 
For any $(\Minf, \go) \in \Mb$, if $H(\Sf, \go)>0$ and 
$H(\Sf, \go)$ does not agree with  $H(\Sf, g)$ identically, 
then for any $\ep > 0$, there exists a 
$(\bar{M}_\infty, \bar{g}_+) \in \Mb$ such that $m(\bar{g}_+) < m(\go)$
and $H(\Sf, \bar{g}_+) \geq H(\Sf, g) - \ep$.
\end{theorem}

Proposition \ref{smthcornerapp} indicates that strict jump of the 
mean curvature at some point on $\Sf$ suggests that there is positive
singular scalar curvature of the piecewise smooth manifold 
$(g, \go)$ at $\Sf$. Thus we expect to level down the singular scalar 
curvature to reduce $m(\go)$. Unlike the proof in 
\cite{Miao}, we must keep the interior geometry of $(\Om, g)$ fixed. 
For that purpose, we first push the 
singular scalar curvature at $\Sf$ into the interior of $\Minf$, 
then we apply conformal 
deformation similar to that in \cite{Miao} outside $\lbar{\Om}$ to 
decrease $m(\go)$. For notation consistency, we let 
$(\lbar{\Om}, \gi)$ denote $(\lbar{\Om}, g)$.

\subsection{A Metric ``Bridge'' near the Boundary}
We establish the existence of a metric ``bridge'' 
that connects $\gi$ and $\go$ near $\Sf$ in a way that the singular scalar 
curvature is propagated into the interior of $\Minf$. 

\begin{prop} \label{bridge}
Assume that $\Sf$ has positive scalar curvature with respect to the induced 
metric and $H(\Sf, \gi) > 0$. For any $(\Minf, \go) \in \Mb$, 
if $H(\Sf, \go) > 0$  
and $H(\Sf, \gi) \geq \neq H(\Sf, \go)$,
then there exists a tubular neighborhood $N_\sig = \Sf \times [0, \sig]$ of 
$\Sf$ in $(\Minf, \go)$ and a scalar flat metric $g_c$ on $N_\sig$ such that
\begin{equation} \label{brgeq}
\left\{
\begin{array}{lcllcl}  
g_c |_{\Sf} & = & \gi |_{\Sf},  & \ \ H(\Sf, g_c) & = & H(\Sf, \gi) \\
g_c |_{\Sfsig} & = & \go |_{\Sfsig}, & \ \ H(\Sfsig, g_c) & \geq \neq & 
H(\Sfsig, \go) ,
\end{array}
\right.
\end{equation}
where ${\Sf}_\sig = \Sf \times \{\sig \}$ and ``$f \geq \neq h$'' means that
``$f \geq h$ but $f$ is not identically $h$''.
\end{prop}

To prove Proposition \ref{bridge}, we adopt the following quasi-spherical 
metric type construction, which was first developed by R. Bartnik in
\cite{Bartnik_spherical} and recently has been used by 
B. Smith and G. Weinstein in \cite{Weinstein} and Y. Shi and L. Tam in
\cite{Shi-Tam}.

Let $\Sf$ be a smooth compact manifold without boundary with dimension $n-1$. 
Let $N = \Sf \times [0, \infty)$ be the product manifold equipped with a 
smooth background metric $g$, which has the form
\begin{equation}
g(x, t) = g_t(x)dx^idx^j + dt^2 ,
\end{equation}
where $t$ is the coordinate on $[0, \infty)$ and $(x^1, x^2, \ldots, 
x^{n-1} )$ are coordinates on $\Sf$. Given a function $\tR$, we want to 
find a function $u > 0$ such that the metric $\tg$ defined by
\begin{equation}
\tg (x, t) = g_t(x)dx^idx^j + u^2(x, t) dt^2
\end{equation}
has the prescribed scalar curvature $\tR$. One basic  
motivation to such a construction is that
\begin{equation} \label{Hfolli}
\tH (x, t) = u(x, t)^{-1} H(x,t) ,
\end{equation}
where $\tH$ and $H$ represent the mean curvature of $\Sf_t = \Sf \times \{t\}$ 
in $(N, \tg)$ and $(N, g)$ with respect to the vector $\pdt$.
The following equation on $u$ was derived 
in several literature (for example, see 
\cite{Bartnik_spherical}, \cite{Shi-Tam}).

\begin{lemma} \label{eqoffolli}
$\tg$ has the scalar curvature $\tR$ if and only $u$ satisfies
\begin{equation} \label{follieq}
H \frac{\partial u}{\partial t} = u^2 \triangle_{g_t} u 
+ \frac{1}{2}(u- u^3)R(g_t) - \frac{1}{2} u R(g) + \frac{1}{2}u^3\tR .
\end{equation}
Here $\triangle_{g_t}(\cdot)$ denotes the Laplacian operator 
of $(\Sf_t, g_t)$, $R(g_t)$ is the scalar curvature of $(\Sf, g_t)$ 
and $R(g)$ is the scalar curvature of $(N, g)$.
\end{lemma}

The following short time existence of solutions follows directly from 
the fact that (\ref{follieq}) is a non-linear parabolic PDE of $u$ 
if $H$ is positive and an implicit function theorem type argument 
(See \cite{Bartnik_spherical}).

\begin{lemma}
For any positive $u_0$ on $\Sf$, there exists a small constant $\sig > 0$ 
and a positive $u=u(x,t)$ on $\Sf \times [0, \sig]$ so that $u$ solves
\begin{equation} \label{ginitialp}
\left\{
\begin{array}{ccl}
H \frac{\partial u}{\partial t}  & = & u^2 \triangle_{g_t} u 
+ \frac{1}{2}(u- u^3)R(g_t)  - \frac{1}{2} u R(g) + \frac{1}{2}u^3\tR \\
u(\cdot, 0) & = & u_0
\end{array}
\right.
\end{equation}
on $N_{\sig} = \Sf \times [0, \sig]$.
\end{lemma}

For our interest in decreasing the mass of $g$ in case $g$ is asymptotically 
flat, we start with $R(g) \geq 0$ and choose $\tR = 0$. 
Then $(\ref{ginitialp})$ is reduced to 
\begin{equation} \label{initialp}
\left\{
\begin{array}{ccl}
H \frac{\partial u}{\partial t} &  = & u^2 \triangle_{g_t} u 
+ \frac{1}{2}(u- u^3)R(g_t) - \frac{1}{2} u R(g) \\
u(\cdot, 0) & = & u_0 .
\end{array}
\right.
\end{equation}
One nice thing about such a choice is that we have a maximum principle 
on the solution to (\ref{initialp}), whose
proof is exactly the same as the proof of the standard maximum principle for 
second order linear parabolic equations.

\begin{lemma} \label{MP}
Assume that the foliation $\{ (\Sf_t, g_t) \}_{t>0} $ has positive 
scalar curvature and positive mean curvature. If $u$ is a positive solution to 
$(\ref{initialp})$ on $N_T = \Sf \times [0, T]$ and $u_0 \leq 1$, 
then $$\max_{N_T} u \leq 1 . $$
\end{lemma}

\noindent {\em Proof of Proposition \ref{bridge}: } We choose 
\begin{equation}
0 < u_0 = \frac{H(\Sf, \go)}{H(\Sf, \gi)} \leq 1
\end{equation}
and let $u$ be a solution to (\ref{initialp}) with $g$ replaced 
by $\go$ on a Gaussian tubular neighborhood 
$N_\sig = \Sf \times [0, \sig]$ of $\Sf$ in $(\Minf, \go)$.
It follows from Lemma \ref{MP} that
$u \leq 1$ on $N_\sig$. Since  $u_0 \not\equiv 1$, by continuity we may 
shrink $\sig$ so that $u(x, \sig) \not\equiv 1$. On $N_\sig$ we define
\begin{equation}
g_c = \got(x) dx^i dx^j + u^2 dt^2 , 
\end{equation}
(\ref{brgeq}) follows directly from (\ref{Hfolli}) and 
Lemma \ref{eqoffolli}. \stop

\subsection{Mass Decrease due to Boundary Effect} \label{mdbi}
We are now in a position to prove Theorem \ref{thdm}. The main idea 
is to first apply 
Proposition \ref{bridge} to propagate the singular scalar curvature at $\Sf$ 
a fixed distance into the interior of $\Minf$,
then to apply Proposition \ref{smthcornerapp} and argument similar to that in
\cite{Miao} to decrease $m(\go)$. 
We divide the proof into several steps. 

\vspace{.2cm}

\noindent \underline{Step 1.} Tilt down the mean curvature to allow a 
strict gap:

\vspace{.1cm}

\noindent For technical reasons, we first approximate $(\Minf, \go)$ by 
$\{ (\Minf, \gos) \}_{s>0}$ where 
$ H(\Sf, \gos) < H(\Sf, \go) $.
Let $\psi$ be a solution to 
\begin{equation}
\left\{
\begin{array}{rcl}
\triangle_{\go} \psi & = & 0 \ \ \ \mathrm{on} \ \Minf \\
\psi & = & 0 \ \ \ \mathrm{on} \ \Sf \\
\psi & \rightarrow & 1 \ \ \ \mathrm{at} \ \infty.
\end{array}
\right.
\end{equation}
For each $s \in (0, 1)$, we define 
\begin{equation}
\gos = ( 1 - s \psi)^\frac{4}{n-2} \go .
\end{equation}
We have that 
\begin{equation} \label{fapp}
\lim_{s \rightarrow 0} m(\gos) = m(\go), \ \ \ \gos |_{\Sf} = \go |_{\Sf}
\end{equation}
and
\begin{equation} \label{Hsg}
H(\Sf, \gos) = H(\Sf, \go) -  (\frac{2s}{n-2}) \frac{\partial \psi}{\pvn} ,
\end{equation}
where $\frac{\partial \psi}{\pvn} > 0$ by the strong maximum principle.
Since $H(\Sf, \go) > 0$, we may assume that
$H(\Sf, \gos) >  \frac{1}{2}H(\Sf, \go) > 0$ 
for sufficiently small $s$ .

\vspace{.2cm}

\noindent \underline{Step 2.} Propagate the singular scalar curvature to the 
interior of $\Minf$:

\vspace{.1cm}

\noindent We apply Proposition \ref{bridge} in a slightly different way in 
order to keep the strict mean curvature gap at $\Sf$. For each
small $s>0$, we let 
\begin{equation}
u^s_0 = \frac{H(\Sf, \gos)}{H(\Sf, \gi) - \frac{2s}{n-2} \frac{\partial \phi}
{\pvn}} = \frac{H(\Sf, \go) - \frac{2s}{n-2} \frac{\partial \phi}{\pvn} }
{H(\Sf, \gi) - \frac{2s}{n-2} \frac{\partial \phi} {\pvn} }
\end{equation}
and let $u_s$ be a short time solution to (\ref{initialp}) with $g$ 
replaced by $\gos$. Since $\gos$ and $u^s_0$ have smooth dependence on $s$,
there exist constants $\sig_0 > 0$ and $s_0 > 0$ so that $u_s$ exists on 
$N_{\sig_0} = \Sf \times [0, \sig_0]$ for any $s \in [0, s_0]$ and $u_s$ 
depends smoothly on $s$. On $N_{\sig_0}$ we define
\begin{equation}
\gsc = \gos(x, t)dx^i dx^j + u_s^2 dt^2
\end{equation} 
where
\begin{equation}
\gos = \gos(x, t)dx^i dx^j + dt^2 .
\end{equation}
It follows from the fact 
$H(\Sf, \gi) > \neq H(\Sf, \go)$ and the proof of Proposition \ref{bridge} 
that $\gsc$ is a scalar flat metric and 
\begin{equation} \label{sbrgeq}
\left\{
\begin{array}{lcllcl}
\gsc |_{\Sf} & = & \gi |_{\Sf},  & \ \ H(\Sf, \gsc) & = & H(\Sf, \gi) - 
\frac{2s}{n-2} \frac{\partial \phi} {\pvn} \\
g^s_c |_{\Sfsigz} & = & \gos |_{\Sfsigz}, & \ \ H(\Sfsigz, \gsc) & \geq & 
(1 + f) H(\Sfsigz, \gos) ,
\end{array}
\right.
\end{equation}
where $0 \leq f \leq 1$ is a function on $\Sfsigz$ that is not identically 
zero and depends only on $\frac{H(\Sf, \go)}{H(\Sf, \gi)}$. We note that,
by choosing $\sig_0$ and $s_0$ sufficiently small, we may also assume that 
$H(\Sfsigz, \gos) > H(\Sfsigz, \go) > \frac{1}{2} H(\Sf, \go) > 0$.

\vspace{.2cm}

\noindent \underline{Step 3.} Smooth $(\gsc, \gos)$ at $\Sfsigz$:

\noindent For each $s \in [0, s_0]$,  
we let  $\tMsinf$ be the modified differentiable manifold on which 
$(\gsc, \gos)$ determines a continuous metric $\gs$ 
as in Section \ref{Mass}. It follows from the proof of 
Proposition 3.1 in \cite{Miao} that
there exists a family of smooth metrics $\{ \gsdel \}_{\del_0 > \del > 0}$ 
on $\tMsinf $, where $\del_0$ is independent on $s$,  such that 
$\gsdel = \gs$ outside $\Sfsigz \times (-\frac{\del}{2}, 
\frac{\del}{2})$, $\gsdel$ approaches to $\gs$ in $C^0$ topology uniformly 
with respect to $s$, and $\Rsdel$, 
the scalar curvature of $\gsdel$, satisfies 
\begin{eqnarray}
  \Rsdel (x,t) &  = & O(1), \ \ for\  (x,t) \in \Sfsigz \times 
  \{\frac{\del^2}{100}
  < |t| \leq \frac{\del}{2}\} \\
  \Rsdel(x,t) &  = & O(1) + \{H(\Sfsigz, \gsc)(x) - H(\Sfsigz, \gos)(x) \} 
  \left\{ 
  \frac{100}{\del^2}\phi(\frac{100t}{\del^2}) \right\}, \nonumber \\
  & & for \  (x,t) \in \Sfsigz \times 
  [-\frac{\del^2}{100},\frac{\del^2}{100}] , 
\end{eqnarray}
where $O(1)$ represents quantity that is bounded 
by constants depending only on $g^0_c$ and $\go $, but not on $\del$ and 
$s$.

\vspace{.2cm}

\noindent \underline{Step 4.} Annihilate the negative scalar curvature:

\noindent For each fixed $s$, we consider the solution to the following 
equation for small $\del$
\begin{equation} \label{bconfu}
\left\{
\begin{array}{rcl}
\triangle_{\gsdel} \usdel + c_n {\Rsdel}_{-} \usdel & = & 0 \ \ 
\mathrm{on\ } \tMsinf \\
\usdel & = & 1 \ \ \mathrm{on\ } \Sf \\
 \usdel & \rightarrow & 1 \ \ \mathrm{at\ } \infty .
\end{array}
\right.
\end{equation}
Similar to Proposition 4.1 in \cite{Miao}, we have that 
\begin{equation} \label{bsupestimate}
\lim_{\del \rightarrow 0} \{ 
\sup_{\tMsinf} \{ | \usdel - 1 | \} \} = 0  
\ \ \mathrm{and} \ \ 
\norm \usdel \norm _{C^{2, \a}(K)} \leq C_K
\end{equation}
for any compact $K \subset \tMsinf \setminus \Sfsigz $. Hence, 
passing to a subsequence, $\usdel$ converges to $1$ on 
$\Sf \times [0, \frac{\sig_0}{2}]$ in $C^2$ topology, which implies that
\begin{equation} \label{bHz}
\lim_{\del \rightarrow 0} \{ \sup_{\Sf} 
 |\frac{\partial \usdel}{\partial \vn}| \} = 0 ,
\end{equation}
where $\vn$ denotes the outward unit normal vector to $\Sf$
determined by $\gsdel$. 
We define 
\begin{equation}
\tgsdel = (\usdel)^{\frac{4}{n-2}} \gsdel ,
\end{equation}
similar to Lemma 4.2 in \cite{Miao}, we have that

\begin{lemma} \label{bmassconverge}
\begin{equation} \label{Hconverge}
\lim_{\del \rightarrow 0} m(\tgsdel) = m(\gos) \ \ \mathrm{and} \ \ 
\lim_{\del \rightarrow 0} H(\Sf, \tgsdel) = H(\Sf, \gsc).
\end{equation}
\end{lemma}

\pf The second limit follows directly from (\ref{bHz}) and
\begin{equation} \label{Htg}
H(\Sf, \tgsdel) = H(\Sf, \gsdel) + \frac{2}{n-2} 
\frac{\partial \usdel}{\partial \vn} .
\end{equation}
To see the first limit, we recall that 
\begin{equation} \label{mb}
m(\tgsdel) = m(\gsdel) + (n-1) A^s_\del ,
\end{equation}
where $A^s_\del$ is given by the expansion
$\usdel = 1 + A^s_\del |x|^{2-n} + O(|x|^{1-n})$.
Applying integration by parts to (\ref{bconfu}) multiplied by $\usdel$, 
we have that
\begin{equation} \label{Ab}
 (2-n)\omega_{n-1} A^s_\del =  \int_{\tMsinf} 
 \left[ | \nabla_{\gsdel} \usdel |^2 - c_n {\Rsdel}_{-} (\usdel)^2 
 \right] d \gsdel + \oint_{\Sf} 
 \frac{\partial \usdel}{\partial \vn} d \mu,
\end{equation}
where $\mu$ is the induced surface measure by $\go$ on $\Sf$.
(\ref{mb}) and (\ref{Ab}) imply that
\begin{equation} \label{bmassdifference}
m(\gsdel) = m(\gsdel) + \frac{n-1}{n-2}\omega^{-1}_{n-1} 
\left\{ \int_{\tMsinf} 
\left[ | \nabla_{\gsdel} \usdel |^2 - c_n {\Rsdel}_{-} {\usdel}^2 \right] 
d \gsdel + \oint_{\Sf} \frac{\partial \usdel}{\partial \vn} d \mu
\right\}   .
\end{equation}
It follows from (\ref{bHz}) and the proof of (\ref{bsupestimate}) 
that the integral term above goes to zero.
Hence, we have that
$ \lim_{\del \rightarrow 0} m(\tgsdel) =
   \lim_{\del \rightarrow 0} m(\gsdel) = m(\gos).$ \stop 

\vspace{.2cm}

\noindent \underline{Step 5.} Level down the positive scalar curvature: 

\noindent To make use of the scalar curvature concentration near $\Sfsigz$ 
as $\del \rightarrow 0$, we let $\vsdel$ be a positive solution to
\begin{equation} \label{bconfv}
\left\{
\begin{array}{rcl}
\triangle_{\tgsdel} \vsdel - c_n {\tRsdel} \vsdel & = & 0 \ \ 
\mathrm{on\ } \tMinf \\
\vsdel & = & 1 \ \ \mathrm{on\ } \Sf \\
\vsdel & \rightarrow & 1 \ \ \mathrm{at\ } \infty, \\
\end{array}
\right.
\end{equation}
and define
\begin{equation}
\hgsdel = (\vsdel)^{\frac{4}{n-2}} \tgsdel.
\end{equation}
Like (\ref{mb}) and (\ref{Ab}), we have that
\begin{equation} \label{mbv}
m(\hgsdel) = m(\tgsdel) + (n-1) A^s_\del ,
\end{equation}
where $A^s_\del$ is given by the expansion
$\vsdel(x) = 1 + A^s_\del |x|^{2-n} + O(|x|^{1-n})$ and can be written 
explicitly as  
\begin{equation} \label{Abv}
 (2-n)\omega_{n-1} A^s_\del =  \int_{\tMsinf} \left[ | \nabla_{\tgsdel} 
 \vsdel |^2  + c_n {\tRsdel} (\vsdel)^2 \right] d \tgsdel + \oint_{\Sf} 
 \frac{\partial \vsdel}{\partial \vn} d \mu .
\end{equation}

\begin{prop} \label{sdelestimate}
\begin{equation} \label{sdelesti}
\liminf_{s \rto 0} \left\{ \liminf_{\del \rto 0} \left\{  \int_{\tMsinf} 
 \left[ | \nabla_{\tgsdel} 
 \vsdel |^2  + c_n {\tRsdel} (\vsdel)^2 \right] d \tgsdel + \oint_{\Sf} 
 \frac{\partial \vsdel}{\partial \vn} d \mu \right\} \right\} > 0 
\end{equation}
\end{prop}

\pf Assume that (\ref{sdelesti}) is not true, then there exist sequences 
$\{ s_k \}$ and $\{ \del_k \}$ so that
\begin{equation} 
\lim_{k \rto \infty} s_k = 0, \ \ \ \lim_{k \rto \infty} \del_k = 0
\end{equation}
and 
\begin{equation} \label{sdelzero}
\lim_{k \rto \infty} \left\{  \int_{\tMkinf} 
 \left[ | \nabla_{\tgkdel} \vkdel |^2  + c_n {\tRkdel} (\vkdel)^2 \right] 
 d \tgkdel + \oint_{\Sf} \frac{\partial \vkdel}{\partial \vn} 
 d \mu \right\} = 0 .
\end{equation}
(\ref{bconfv}) implies that, passing to a subsequence, 
$\{ \vkdel \}$ converges to a 
$g^0_c$-harmonic function $v$ on the compact set 
$\Sf \times [0, \frac{\sig_0}{4}]$ in $C^2$ topology, where 
$0 \leq v \leq 1$ by the maximum principle. Hence,
\begin{equation} \label{Hvkdelv}
\deltoz \{ \sup_{\Sf} \{ | \frac{\partial \vkdel}{\partial \vn} - 
\frac{\partial v}{\partial \vn} | \} \} = 0 .
\end{equation}
We claim that $v \equiv 1$. If not, the strong maximum principle implies 
that 
\begin{equation} \label{pnvsupv1}
\sup_{\Sf} \frac{\partial v }{\partial \vn} < 0 \ \ \mathrm{and} \ \ 
0 < v(x) < 1 \  \mathrm{for} \ x \in \Sf \times (0, \frac{\sig_0}{4}) .
\end{equation}
We let $\theta \in (0, 1)$  denote the supremum of $v$ on 
$\Sf \times \{ \frac{\sig_0}{8} \} $ and let $w_k$ be the solution to 
\begin{equation}\label{bbarrier}
\left\{ 
\begin{array}{rl}
\triangle_{\tgkdel} w_k = 0 & \mathrm{\ on\  } \tMkinf \setminus 
N_{\frac{\sig_0}{8}} \\
w_k = \theta & \ \mathrm{on} \ \Sf_\frac{\sig_0}{8} \\
w_k (x) \rightarrow 1  & \mathrm{\ at\  } \infty ,
\end{array} 
\right.
\end{equation}
where $N_\frac{\sig_0}{8}= \Sf \times [0, \frac{\sig_0}{8}]$. 
It follows from (\ref{bconfv}), (\ref{bbarrier}) and the maximum 
principle that $w_k \geq \vkdel$, which implies that
\begin{equation} \label{ABdel}
A_{\del_k} \leq B_k ,
\end{equation}
where $B_k$ is given by the expansion $w_k  = 1 + B_k |x|^{2-n} 
+ O(|x|^{1-n})$ and can be written explicitly as  
\begin{equation} \label{Bbw}
 (2-n) \omega_{n-1} B_k = \int_{\tMkinf \setminus 
 N_\frac{\sig_0}{8} } | \nabla_{\tgkdel} w_k |^2  d \tgkdel 
 + \oint_{\Sf_{\frac{\sig_0}{8}}} \frac{\partial w_k}{\partial \vn} \ 
 d (\tgkdel |_{\Sf_\frac{\sig_0}{8}}) .
\end{equation}
By the maximum principle, we have that $ \frac{\partial w_k}{\partial \vn} 
\geq 0 $. Hence, (\ref{Bbw}) shows that
\begin{equation} \label{Bbwest}
(2-n) \omega_{n-1} B_k \geq  \int_{\tMkinf \setminus 
N_{\frac{\sig_0}{8}} } | \nabla_{\tgkdel} w_k |^2  d \tgkdel .
\end{equation}
Since $\theta \in (0, 1)$, (\ref{bbarrier}) implies that 
\begin{equation} \label{Bbound}
\lim_{k \rto \infty} \int_{\tMkinf \setminus N_\frac{\sig_0}{8} } 
| \nabla_{\tgkdel} w_k |^2  d \tgkdel \geq \frac{1}{2} E(g, \theta) > 0 ,
\end{equation}
where $E(g, \theta) = \inf \{ \int_{\tilde{M}^0_\infty 
\setminus N_\frac{\sig_0}{8}} | \nabla_{g_0} \psi |^2 d g_0
\ | \ \psi = \theta \ on\ \Sf_\frac{\sig_0}{8}, \ \psi \rightarrow 1 \ 
at \ \infty \}$.
Thus it follows from (\ref{ABdel}), (\ref{Bbwest}) and (\ref{Bbound}) that
\begin{equation} 
\lim_{k \rto \infty} (2-n)\omega_{n-1} 
A_{\del_k} \geq \frac{1}{2}E(g, \theta) .
\end{equation}
By (\ref{Abv}) we have a contradiction to (\ref{sdelzero}) and 
$v \equiv 1$ on $\Sf \times [0, \frac{\sig_0}{8}]$. Now
(\ref{Hvkdelv}) imply that
\begin{equation}
\lim_{k \rto \infty} \{ \sup_{\Sf} | \frac{\partial \vkdel}{\pvn} | \} = 0.
\end{equation}
Thus (\ref{sdelzero}) is reduced to 
\begin{equation} \label{similar}
\lim_{k \rto \infty} \int_{\tMkinf} 
 \left[ | \nabla_{\tgkdel} \vkdel |^2  + c_n {\tRkdel} (\vkdel)^2 \right] 
 d \tgkdel = 0 .
\end{equation}
Now we are in a situation that is as same as in Proposition 4.2 
in \cite{Miao}. The proof in \cite{Miao} shows that (\ref{similar}) 
can not hold. \stop

\vspace{.2cm}

To complete the proof of Theorem \ref{thdm}, we fix a constant $\ep>0$ and 
define 
\begin{equation}
\hat{g}^s_{\del t} = (1 + t(\vsdel - 1))^\frac{4}{n-2} \tgsdel .
\end{equation}
The mean curvature relation 
\begin{equation}
H(\Sf, \hat{g}^s_{\del t}) = H(\Sf, \tgsdel) + 
t \frac{\partial \vsdel}{\pvn} < H(\Sf, \tgsdel)
\end{equation}
and the fact that, passing to a subsequence, $\{\vsdel \}$ converges to a 
$g^0_c$-harmonic function $v$ on $\Sf \times [0, \frac{\sig_0}{8}]$
in $C^2$ topology as $\del, s \rightarrow 0$ imply that there exists a constant
$t_0 > 0$ depending only on $\ep, v$ but not on $\del, s$ such that
\begin{equation}
H(\Sf, \tgsdel) - \12 \ep < H(\Sf, \hat{g}^s_{\del t_0}) < H(\Sf, \tgsdel) .
\end{equation}
Now it follows from (\ref{fapp}), (\ref{sbrgeq}), Lemma \ref{bmassconverge} 
and Proposition \ref{sdelestimate} that
there exist $0<s<s_0$ and $0< \del < \del_0$ such that
\begin{equation}
\left\{
\begin{array}{ccc}
m(\gos) & < & m(\go) + \frac{t_0}{4} A \\
H(\Sf, \gsc) & = & H(\Sf, \gi) - \frac{2s}{n-2} 
\frac{\partial \psi}{\pvn} \\
m(\tgsdel) & < & m(\gos) + \frac{t_0}{4} A \\
H(\Sf, \tgsdel) & < & H(\Sf, \gsc) + \frac{s}{n-2} 
\frac{\partial \psi}{\pvn} \\
H(\Sf, \tgsdel) & > & H(\Sf, \gsc) - \frac{s}{n-2} 
\frac{\partial \psi}{\pvn} \\
m(\hgsdeltz) & \leq & m(\tgsdel) - \frac{t_0}{2} A ,
\end{array}
\right.
\end{equation} 
where $|\frac{s}{n-2} \frac{\partial \psi}{\pvn}| < \frac{\ep}{8}$ and
$$A = \liminf_{s \rto 0} \left\{ \liminf_{\del \rto 0} \left\{  \int_{\tMsinf} 
 \left[ | \nabla_{\tgsdel} 
 \vsdel |^2  + c_n {\tRsdel} {\vsdel}^2 \right] d \tgsdel + \oint_{\Sf} 
 \frac{\partial \vsdel}{\partial \vn} d {\gos}_\Sf  \right\} \right\} > 0 . $$ 
Hence, we have $(\tMsinf, \hgsdeltz) \in \Mb$ such that 
\begin{equation}
\left\{
\begin{array}{ccc}
m(\hgsdeltz) & < & m(\go) \\ 
H(\Sf, \hat{g}^s_{\del t_0}) & > & H(\Sf, \gi) - \ep .
\end{array}
\right.
\end{equation} 
\stop

\subsection{A Note on Quasi-local Mass} \label{note}
Theorem \ref{thdm} suggests that, when considering the quasi-local mass 
question, we may first focus on a domain $(\Om, g) \subset (M^3, g)$ 
which has quasi-convex boundary, i.e. 
$\partial \Om$ has positive scalar curvature and positive mean curvature. 
For such a domain $(\Om, g)$, we modify Bartnik's definition slightly to 
define the following metric extension class 
\begin{eqnarray*} 
\lPM & = & \{ (\Minf, \go) \ | \ (\Minf, \go)\ is\ 
a\ smooth\ asymptotically\ flat\ manifold \nonumber \\
& & \hspace{2cm} with\ boundary\ \Sf \ such\ that\ R(\go) \geq 0, \  
g|_{\Sf} = \go |_{\Sf}, \nonumber \\
& & \hspace{1.8cm} \ 
\ H(\Sf, g) \geq H(\Sf, \go) > 0, \ and\ there\ is\ no\ \mathbb{S}^2\  
\nonumber \\
& & \hspace{2cm} outside\ \Sf\ with\ area\ less\ than\ or\ equal\ to\
\Sf. \} .
\end{eqnarray*}
Like the no horizon assumption in Bartnik's definition, the last 
restriction on $\lPM$ is imposed to prevent the infimum of the total 
mass functional over $\lPM$ from being trivially zero.
A careful examination of our construction in Section \ref{mdbi} 
reveals that, if we start with a $(\Minf, \go)$ in $\lPM$,
the resulting comparison manifold $(\tMsinf, \hgsdeltz)$ can also be 
chosen in $\lPM$. Hence, we have the following characterization of $\lPM$.

\begin{prop} \label{bdryinf}
For any given $\ep > 0$, 
\begin{equation*} 
\inf \{ m(\go) \ | \ \go \in \lPM \}
= \inf \{ m(\go) \ | \ \go \in \lPM, \ H(\Sf, \go) \geq 
H(\Sf, g) - \ep \} .
\end{equation*} 
\end{prop}

Proposition \ref{bdryinf} implies that, for any mass minimizing sequence 
$\{(M_{\infty_i}, g_{+_i})\}_i$ in $\lPM$, it can always be 
replaced by a new mass minimizing sequence $\{(\tM_{\infty i}, \tg_{+_i})\}_i$
such that $\{ H(\Sf, \tg_{+i}) \}_i$ monotonically increases to $H(\Sf, \gi)$. 
In particular, we may assume that the Hawking mass of $\Sf$ in 
$(\tM_{\infty i}, \tg_{+ i})$
\begin{equation}
\frac{Area(\Sf)^\12}{(16\pi)^\frac{3}{2}}
\left\{ 16 \pi - \int_{\Sf}  H(\Sf, \tg_{+i})^2 \right\}
\end{equation}
monotonically decrease to the Hawking mass of $\Sf$ in $(\lbar{\Om}, \gi)$. 

We call a manifold $(\Minf, g_{min}) \in \lPM$ a 
{\bf minimal mass extension} of $(\lbar{\Om}, g)$ if 
$$ m(g_{min}) = \inf \{ m(\go) \ | \  (\Minf, \go) \in \lPM \} . $$
So far, it is an open question if there exists a minimal mass extension
of $(\lbar{\Om}, g)$ in $\lPM$. 
However, the following proposition shows that if such an extension exists,
it must be a static metric \cite{Corvino} with zero scalar curvature and 
satisfy Bartnik's geometric boundary condition \cite{Bartnik_energy}. 
For notation consistency, we again let $\gi$ denote $g$ on $\lbar{\Om}$.

\begin{prop} \label{minisflat}
If $(\Minf, \go)$ is a minimal mass extension of $(\lbar{\Om}, \gi)$, 
then, in the interior of $\Minf$, $\go$ is a scalar flat and static metric
and, at the boundary $\Sf$, $\go$ satisfies 
\begin{equation} \label{bdH}
\left\{
\begin{array} {ccc}
\gi |_\Sf & = & \go |_{\Sf} \\ 
H(\Sf, \gi) &  =  & H(\Sf, \go) .
\end{array}
\right.
\end{equation}
\end{prop}

\begin{rmk}
For a great introduction to static metrics, readers are
referred to \cite{Corvino}.
\end{rmk}

\pf The boundary condition (\ref{bdH}) follows directly from 
Theorem \ref{thdm}. To derive the interior equation, we first show that
$R(\go)$ is identically zero. Assume not, then there exists a positive 
solution $u$ to 
\begin{equation}\label{eqofu}
\left\{ \begin{array}{rcll}
        \triangle_{\go} u - c_n R(\go) u & = & 0 & on\ \Minf \\
         u & = & 1 \ & on\ \Sf \\
         u & \rightarrow & 1 & at\ \infty \ ,
        \end{array} 
\right.  
\end{equation} 
and $u$ has an asymptotic expansion at $\infty$,
\begin{equation}
u(x) = 1 + \frac{A}{|x|^{n-2}} + O(|x|^{1-n}).
\end{equation}
It follows from the strong maximum principle that $A < 0$. 
We consider a path of metrics $\{ \got \}_{0 \leq t \leq 1}$ defined by 
$ \got = v_t^\frac{4}{n-2} \go, $
where 
\begin{equation} \label{defvt}
 v_t = (1-t) + t u = 1 + t(u - 1 ).
\end{equation}
It follows from (\ref{eqofu}) that
\begin{equation}
\triangle_{\go} v_t - c_n R(\go) v_t =  
         c_n(t-1)R(\go) \leq 0 .
\end{equation}
Hence, $(\Minf, \got)$ is asymptotically flat with non-negative scalar 
curvature.
At $\Sf$, we have that 
\begin{equation} \label{hcurvature}
\left\{
\begin{array} {ccl}
 \got |_{\Sf} & = & \go |_{\Sf} \\ 
H(\Sf, \got) & = & H(\Sf, \go) + \frac{2}{n-2}\frac{\partial v_t}{\pvn}, \\
\end{array}
\right.
\end{equation}
where $\vn$ is the outward unit normal vector field along $\Sf$ determined 
by $\go$. By the strong maximum principle,  
$\frac{\partial u}{\pvn} < 0 $ at every point on $\Sf$. 
Hence, it follows from (\ref{defvt}) and (\ref{hcurvature}) that,
for sufficiently small $t_0$, $\{ \got \}_{0 \leq t \leq t_0} \subset \lPM$.
On the other hand, straightforward calculation reveals that
\begin{equation}
m(\got) = m(\go) + (n-1)tA.
\end{equation}
The fact $A < 0$ implies that $m(\got) < m(\go)$ for small positive $t$, which 
contradicts the fact $(\Minf, \go)$ minimizes the total mass.
Hence, $\go$ must have vanishing scalar curvature on $\Minf$. 

Now assume that $\go$ is not static on $\Minf$, the scalar curvature 
deformation result of J. Corvino \cite{Corvino} then implies that 
there exists a manifold $(\Minf, \tg) \in \lPM$ such that $\tg$ agrees 
with $g$ outside a compact set $K \subset ( \Minf \setminus \Sf )$ 
and $R(\tg) > 0$ on $K$. Thus $(\Minf, \tg)$ is a minimal mass extension 
with non-zero scalar curvature,
which is a contradiction to what we have just proved. Hence, $\go$ is 
a static metric. \stop

\vspace{.2cm}

Proposition \ref{minisflat} suggests one interesting metric extension question 
known as the static metric extension conjecture proposed by R. Bartnik in 
\cite{Bartnik_energy}. For a partial answer to the corresponding small 
data solution, readers are referred to \cite{Miao_static}.

\bibliographystyle{plain}
\bibliography{bdrymass}

\end{document}